\documentclass[epj,nopacs]{svjour}

\usepackage{amsmath,amssymb,graphicx,bm}
\usepackage{color}
\usepackage{multicol}        
\usepackage[bottom]{footmisc}

\newcommand{\beq}{\begin{equation}}
\newcommand{\eeq}{\end{equation}}
\newcommand{\bea}{\begin{eqnarray}}
\newcommand{\eea}{\end{eqnarray}}
\newcommand{\ba}{\begin{align}}
\newcommand{\ea}{\end{align}}
\newcommand{\bfig}{\begin{figure}}
\newcommand{\efig}{\end{figure}}

\newcommand{\D}{\displaystyle}
\newcommand{\rg}{\raggedright}
\newcommand{\gev}{\, \text{GeV}}
\newcommand{\mev}{\, \text{MeV}}
\newcommand{\thetaprime}{\theta^{\prime}}
\newcommand{\thetain}{\theta_{\rm in}}
\newcommand{\tpi}{t_{\pi}}
\newcommand{\tin}{t_{\rm in}}
\newcommand{\la}{\langle}
\newcommand{\ra}{\rangle}
\newcommand{\piprime}{\Pi^{\,\prime}}

\begin{document}


%
\title{New constraints on the Pion EM form factor using $\piprime(-Q^2)$}
\author{Gauhar Abbas \and B.\ Ananthanarayan \and S. Ramanan}

\institute{Centre for High Energy Physics,
Indian Institute of Science, Bangalore\ 560 012, India.\\
{\email{gabbas@cts.iisc.ernet.in} \\
 \email{anant@cts.iisc.ernet.in} \\ 
 \email{suna@cts.iisc.ernet.in}}}        

\date{\today}
%
\abstract{We study the constraints arising on the expansion
parameters $c$ and $d$ of the Pion electromagnetic form factor
from the inclusion of pure space-like data and the phase of time-like
data along with one space-like datum, using as input the first derivative of the 
QCD polarization amplitude $\piprime(-Q^2)$.  These constraints when
combined with other analyses, provide a valuable check on a
determination of $c$ due to Guo et al.
and on our previous work where pionic contribution to the 
$(g-2)$ of the muon was used as the input. This work further illustrates the power of analyticity techniques in form factor analysis.
}
\titlerunning{Pion EM form factor and $\piprime(-Q^2)$}
\authorrunning{Gauhar Abbas \and B. Ananthanarayan \and S.Ramanan} 
\maketitle

\section{Introduction}
\label{introduction}

The Pion form factor continues to be of current 
interest~\cite{leutwyler,AR1,Masjuan,Raha,Boyle,AR2}. 
In~\cite{AR1,AR2}, we developed a framework for obtaining 
constraints on the low-energy expansion coefficients $c$ and $d$ of the 
Pion form factor using data from the space-like region 
($t < 0$)~\cite{Bebek1,Brown,Bebek2,Amendolia,nucl-ex/0607007}, 
and the simultaneous inclusion of the phase of time-like data and
one space-like datum respectively, where  
$c$ and $d$ are the Taylor coefficients in the low-energy 
expansion of the Pion EM form factor given as,
\beq
	F_\pi(t) = F_\pi(0) + \D\frac{1}{6} \la r^2_\pi \ra t + c t^2 + d t^3 + \cdots.
	\label{ffexp:Eq}
\eeq
In~\cite{AR2}, we used as
input the pionic contribution to the $(g-2)$ of the muon.
This technique can be fruitfully extended to other inputs.

Our work has been greatly inspired by the results of
Caprini~\cite{Caprini1} who has studied in great detail
the problem of obtaining allowed regions in the $c$-$d$ plane using
the first derivative $\piprime(q^2)$ of the QCD vacuum polarization amplitude $\Pi(q^2)$, which satisfies the following dispersion relation, 
\beq
	\piprime(q^2) = \D\frac{1}{\pi} \int_0^{\infty} \frac{\text{Im} \Pi(t + i\epsilon)}{(t - q^2)^2} \,dt,
	\label{eqn:disp1}
\eeq
first with no constraints from data, followed by the inclusion of the phase of the form factor up to an energy
denoted by $t_{in}$.  
Caprini has shown that the modulus of the form factor in
the time-like region can further improve the bounds, via  
an extensive construction of a function
that is analytic in the cut plane, where the cut begins at $t_{in}$.

In the present work, we wish to apply the methods developed in
Ref.~\cite{AR1,AR2} to $\piprime(q^2)$. Firstly, it was pointed
out in Ref.~\cite{AR2}, that if 
$c$ is taken to lie in the range $4.49\pm 0.28\gev^{-4}$, a value presented
in Ref.~\cite{BT} based on chiral perturbation theory (for
important early work on the Pion electromagnetic form factor
in chiral perturbation theory see Ref.~\cite{Gasser:1990bv}),
a possible conflict arises between the regions isolated in our work
using the phase of time-like data and one space-like datum where
$(g-2)$ is taken as the input, and that of Caprini that uses magnitude and
phase of time-like data and $\piprime(q^2)$ as input. Therefore it is
contingent to check whether such a conflict also arises
if we were to use $\piprime(q^2)$ itself as the input. One of
the objectives of this work is to explore this issue.  Secondly,
a very recent determination of $c$ yields the value $4.00$ $\pm$ $0.50$ 
$\gev^{-4}$~\cite{Guo:2008nc}.  We must therefore ask what the status of the 
purported discrepancy is for this set of values.  It turns out that
the discrepancy is substantially mitigated for values of c in the range $(3.5 \gev^{-4}$ - $4.0 \gev^{-4})$.  In fact with the earlier
determination of the same quantity 
$3.2\pm 0.5 \pm 0.9\gev^{-4}$~\cite{hep-ph/9604279}, the central
value is accommodated in the overlap region of the ellipses we
find in the present work, and the one isolated by Caprini with
modulus as well as phase of time-like data.  In particular, one
may conclude from here that the value of $d$ is typically about
$10 \gev^{-6}$ for $c \sim 3.2 \gev^{-4}$ and about $20 \gev^{-6}$ for $c \sim 4.0 \gev^{-4}$. This is one of the important conclusions of this work.

This paper is organized as follows. In Section~\ref{sect:formalism} we provide a general discussion on the origin
of bounds on which the current analysis is based and present an extensive discussion of the formalism.
This is followed by a discussion on the inclusion of space-like data 
and our results in Section~\ref{sect:spacelike}.  Section~\ref{sect:tl_sl} discusses the 
inclusion of the phase of time-like data and one space-like datum and
our results.  We then present our final conclusions and a discussion in Section~\ref{sect:conclusion}.

\section{Formalism}
\label{sect:formalism}


The Pion electromagnetic form factor $F_\pi(t)$ enters 
several observable quantities through expressions of the type:
\beq
\D\frac{1}{\pi} \int_{4 m_\pi^2}^\infty dt\, \rho(t) |F_\pi(t)|^2
\label{eqn2:bound}
\eeq
where $\rho(t)>0$ in the region of integration. Consider as an example the first derivative of the QCD vacuum polarization $\Pi(-Q^2)$ evaluated at a space-like value
of $q^2 = -Q^2$, which satisfies the dispersion relation given in Eq.~\ref{eqn:disp1} and hence considering only the Pion contribution as in~\cite{Caprini1}, $\piprime(-Q^2)$ satisfies the following inequality:
\beq
\piprime(-Q^2) \ge \D\frac{1}{48 \pi^2} \int_{t_\pi}^\infty
\frac{dt}{(t+Q^2)^2} \left(1-\frac{t_\pi}{t}\right)^{3/2} |F_\pi(t)|^2,
\label{ineq:Eq}
\eeq
where $t_\pi = 4 m_\pi^2$.
Since this expression has a positive-definite integrand, it would
be of interest to find a lower bound for it.  

This objective
is achieved by first mapping the form factor in the $t$ plane to the $z$ plane through the following conformal map:
\beq
\D \frac{z+1}{z-1} = -i \sqrt{\frac{t - t_\pi}{t_\pi}}
\label{ztmap.eqn1}
\eeq 
resulting in a unit circle and the branch cut lies at its circumference. As a result $F_\pi(z)$ is analytic within the unit circle in the $z$ plane.
Note that this definition is in accordance with the convention
of Caprini~\cite{Caprini1} and is related to the map defined in~\cite{AR1,AR2} through $z \rightarrow -z$. As a result of this transformation, Eq.~\ref{ineq:Eq} becomes,
\beq
I \ge \D\frac{1}{2 \pi} \int_0^{2 \pi} d\theta |h(\exp{i\theta})|^2,
\label{bound:Eq}
\eeq
where
\beq
 h(z) = f(z) w_\pi(z)
\label{hz.eqn1} 
\eeq
and $z = \exp(i \theta)$.
Here, $I$ is the value of $\piprime(-Q^2)$, $f(z)$ is the form factor in terms of the conformal variable and $w_\pi(z)$ is the ``outer function'' to be discussed further.
Within the unit circle, $f(z)$ is analytic, therefore, the function $h(z)$ admits an expansion given by
\begin{equation}
h(z)=a_0+a_1 z+ a_2 z^2 + \cdots,
\label{hz:Eq}
\end{equation}
where the $a_n$ are real and are functions of the expansion coefficients $c$ and $d$ of the Pion EM form factor given in Eq.~\ref{ffexp:Eq}.
From the Parseval theorem of Fourier analysis, the integral in Eq.~\ref{bound:Eq} is now simply given by,
\beq
	I \ge \sum_{n = 0}^{\infty} a_n^2.
\label{cons.eqn2}
\eeq
Due to the fact that this quantity is a sum of squares, truncating
the series yields a lower bound for $I$.

In the past, the observable that was studied in great detail
was the pionic contribution to the muon anomalous magnetic moment $(g-2)$, i.e, $I$ was the bound on the $(g-2)$ of the muon, 
where the normalization and the charge radius of
the Pion were supplied from experiments, and the series was truncated 
after the first two terms~\cite{Palmer}. A corresponding outer function for the relevant dispersion relation gives rise to a set of coefficients $a_n$ with $n = 0, 1, \cdots$.
One could have improved the bound $I$ by retaining the expression up to the three
terms shown above, and supplying in addition 
the second Taylor coefficient
coefficient ($c$).  If we were to retain the series up to the first
four terms, we would have had to supply the third Taylor
coefficient ($d$) as well. 
Improvements on the bounds were made possible by using
experimental information on the Pion form factor at space-like~\cite{RainaSingh} and at time-like~\cite{Raszillier:1975wu} values
of the momentum transfer.

In this work, as in~\cite{AR1,AR2}, we ask the reverse question: given the value for $I$, which in this case is the value of $\piprime(-Q^2)$ and the available data in both space-like and time-like regions, can one obtain bounds on the Taylor coefficients $c$ and $d$ of the Pion form factor?  

In Ref.~\cite{AR1}, we considered only the inclusion of pure
space-like data taking up to three constraints. Our results
were encouraging and provided constraints that were significantly
stronger than those found by Caprini~\cite{Caprini1} using
only phase of time-like data and the value $\piprime(-Q^2)$.

In Ref.~\cite{AR2},
we considered the problem of wiring in the constraints from the
phase of the form factor in the time-like region and the constraint
from one space-like datum.  The corresponding Lagrange multiplier
technique was developed and applied. One significant finding was that 
the simultaneous inclusion of these constraints isolated the
allowed region as one that was significantly smaller than the mere
intersection of the constraints taken one at a time.  
In~\cite{AR2}, we have also carried out a detailed analysis
with several alternative models for the phase of the form factor,
using the simple scattering phase formula, two Roy equation fits
which, via the Fermi-Watson theorem, get related to the phase of
the form factor, and also directly from the parametrization of
the form factors themselves.

Specializing to $\piprime(-Q^2)$, which obeys the inequality in Eq.~\ref{ineq:Eq}, the map from the $t$-plane to the $z$-plane results in the following closed expression for the outer function  $w_\pi(z)$,
\beq
w_\pi(z) = \D\frac{(1-d_\pi)^2}{16} \sqrt{\frac{1}{6 \pi t_\pi}}
\frac{(1+z)^2 \sqrt{1-z}}{(1-z d_\pi)^2},
\label{w_pi:Eq}
\eeq
where,
\beq
d_\pi=\D\frac{\sqrt{t_\pi+Q^2}-\sqrt{t_\pi}}{\sqrt{t_\pi+Q^2}+\sqrt{t_\pi}}.
\eeq

Retaining only a finite number of terms $N = 3$ in Eq.~\ref{hz:Eq}, the expansion coefficients for
the function $h(z)$ in terms of $f(z)$ and $w_{\pi}(z)$
may be readily expressed as follows, using the expansion for the Pion form factor (Eq.~\ref{ffexp:Eq}),
\beq
a_0 = h(0) = w_\pi(0),
\eeq
\beq
a_1 = h^{\prime}(0)= w_\pi^{\prime}(0) -\D\frac{2}{3} r_\pi^2 t_\pi w_\pi(0),
\eeq
\bea
a_2 & = &\D\frac{h^{\prime \prime}(0)}{2!}=\frac{1}{2}\left[
	w_\pi(0)\left(-\frac{8}{3} r_\pi^2 t_\pi + 32 \,c \, t_\pi^2\right) \right] \nonumber \\ 
	 &+& \frac{1}{2} \left[2 w_\pi^{\prime}(0)\left(-\frac{2}{3} r_\pi^2 t_\pi\right) + w_\pi^{\prime \prime}(0) \right], 
\eea
and
\bea
a_3 &=& \D\frac{h^{\prime \prime \prime}(0)}{3!} =
\D\frac{1}{6} \left[w_\pi(0)\left(- 12 r_\pi^2 t_\pi + 384\, c\, t_\pi^2 - 384 \,d \, t_\pi^3\right) \right] \nonumber \\
&+& \D\frac{1}{6}\left[3 w_\pi^{\prime}(0)\left(-\frac{8}{3} r_\pi^2 t_\pi + 32 \,c \, t_\pi^2\right)\right] \nonumber \\
&+&\D\frac{1}{6} \left[
-\,2 w_\pi^{\prime \prime}(0) r_\pi^2 t_\pi + w_\pi^{\prime \prime \prime}(0) \right].  
\eea

Scaling Eq.~(\ref{cons.eqn2}) by $I$ and keeping only $N$ terms, we get
\beq
	\mu_0^2 = \sum_{n = 0}^{N} (c_n)^2 \le 1,
	\label{cons.eqn3}
\eeq
where,
\beq
	c_n = \D\frac{a_n}{\sqrt{I}}.
	\label{cap_coeffn}
\eeq
Any $c_n$ which satisfies $\mu_0^2 \le 1$ is allowed and the 
equality gives the bound. We have already seen that this yields an 
ellipse in the $c-d$ plane.

The bound $I$ is evaluated at $Q^2= 2 \gev^2$ so that $I = 0.009546\gev^{-2}$ and $r_\pi^2 = 0.42\, {\rm fm}^2$, as in Ref.~\cite{Caprini1}, in order to have a meaningful
comparison.
Including up to the second (third) derivative 
for $F_{\pi}(t)$ results in constraints for $c$ ($c$ and $d$). 

\section{Space-like constraints}
\label{sect:spacelike}

Additional constraints in the $c-d$ plane are now obtained by
wiring in information coming from the space-like region, which can be expressed as linear constraints:
\beq
	h(x) - \sum_{n = 0}^{\infty} a_n x^n = 0,
	\label{eq:spacelike}
\eeq
where $h(x)$ includes the value of the form factor $F_\pi(t)$ at space-like points defined by Eq.~\ref{hz.eqn1} and $x$ is real, lying in the range $(0 < x < 1)$ as a result of the conformal map. The constraints are included through the method of Lagrange multipliers. We set up the following Lagrangian:
\beq
L= \D\frac{1}{2} \sum_{n=0}^\infty
 c_n^2 +\sum_{m=1}^M \alpha_m (J_m - \sum_{n=0}^\infty c_n z^n_{m}),
\label{space_cons.eqn2}
\eeq
where $J_m = h(x_m)/\sqrt{I}$, $h(x_m)$ are the space-like constraints defined at $z_m = x_m,\, m = 1,2,3 \cdots$. 
In Eq.~(\ref{space_cons.eqn2}), we wish to consider only finite number of expansion coefficients $c_n$ and therefore we set $N = 3$.
With the space-like constraints included, the equation satisfied by the coefficients $c_n$ (analogous to Eq.~\ref{cons.eqn3}) is obtained by eliminating the Lagrange multipliers $\alpha_m$ and can be written conveniently in the form of the following determinantal equation,
\beq
\left|
\begin{array}{c c c c c c c c}
1 & c_0 & c_1 & c_2 & c_3 & \D\frac{h(x_1)}{\sqrt{I}} & \D\frac{h(x_2)}{\sqrt{I}} & \cdots\\
c_0 & 1 & 0 & 0 & 0 & 1 & 1 & \cdots\\
c_1 & 0 & 1 & 0 & 0 & x_1 & x_2 & \cdots \\
c_2 & 0 & 0 & 1 & 0 & x_1^2 & x_2^2 & \cdots\\
c_3 & 0 & 0 & 0 & 1 & x_1^3 & x_2^3 & \cdots \\
\D\frac{h(x_1)}{\sqrt{I}} & 1 & x_1 & x_1^2 & x_1^3 & (1-x_1^2)^{-1} & (1-x_1 x_2)^{-1} & \cdots\\
\D\frac{h(x_2)}{\sqrt{I}} & 1 & x_2 & x_2^2 & x_2^3 & (1-x_2 x_1)^{-1} & (1-x_2^2)^{-1} & \cdots\\
\vdots & \vdots & \vdots & \vdots & \vdots & \vdots & \vdots & \\
\end{array}\right|=0.
\label{det.eqn1}
\eeq
Note that we use the scaled coefficients $c_n$ as opposed to $a_n$ in Ref.~\cite{AR1} for the sake of notational convenience and as a result we need to scale $h(x_m)$.
Solving for $c$ and $d$ gives an ellipse in the $c-d$ plane, when $I$ is used as an input.
  
The data we
use are given in
Tables~\ref{table_bebek},~\ref{table_jlab} and~\ref{table_amen} that list the values of $x(t)$, the space-like points for a given $t$ and the 
corresponding value of $F_\pi(t)$. The value of $h(x)$ can be evaluated using Eqs.~\ref{hz.eqn1} and~\ref{w_pi:Eq}.
To make our notations clear, we refer to the data points corresponding 
to a particular $|t|$ as $x_1$, $x_2$ and so on, 
in the ascending order of magnitude of $|t|$. 
The tables also show the experimental errors in the data. 

\begin{table*}
	\begin{center}
	\caption{Spacelike data from Bebek et.al~\cite{Bebek2}}
	\label{table_bebek}
		\begin{tabular}{llll}
		\hline\noalign{\smallskip}
		 & $t$($-Q^2$) [$\gev^2$] & $F_{\pi}(t)$ & $x(t)$   \\
		\noalign{\smallskip}\hline\noalign{\smallskip}     	
		1 & -0.620 & 0.453 $\pm$ 0.014 &  0.499 \\ 
		2 & -1.216 & 0.292 $\pm$ 0.026 &  0.606 \\ 
		3 & -1.712 & 0.246 $\pm$ 0.017 &  0.655 \\ 
		\noalign{\smallskip}\hline
	\end{tabular} 
	\end{center}
	\vspace*{0.2cm}
	\begin{center}
	\caption{Spacelike data from Tadevosyan et al.~\cite{nucl-ex/0607007}}
	\label{table_jlab}
		\begin{tabular}{llll}
		\hline\noalign{\smallskip}
		 & $t$($-Q^2$) [$\gev^2$] & $F_{\pi}(t)$ & $x(t)$  \\
		\noalign{\smallskip}\hline\noalign{\smallskip}    	
		1 & -0.600 & 0.433 $\pm$ 0.017 &  0.494 \\ 
		2 & -1.000 & 0.312 $\pm$ 0.016 &  0.576 \\
		3 & -1.600 & 0.233 $\pm$ 0.014 &  0.645 \\  
		\noalign{\smallskip}\hline
		\end{tabular}
	\end{center}
	\vspace*{0.2cm}
	\begin{center}
	\caption{Spacelike data from Amendolia et.al~\cite{Amendolia}}
	\label{table_amen}
		\begin{tabular}{llll}
		\hline\noalign{\smallskip}
		 & $t$($-Q^2$) [$\gev^2$] & $F_{\pi}(t)$ & $x(t)$  \\
		\noalign{\smallskip}\hline\noalign{\smallskip}  	
		1 & -0.131 & 0.807 $\pm$ 0.015 &  0.242\\
		2 & -0.163 & 0.750 $\pm$ 0.016 &  0.275\\ 
		\noalign{\smallskip}\hline
		\end{tabular}
	\end{center}
\end{table*}

The data from various experiments are chosen so as to give reliable numerical 
results. 
As in Ref.~\cite{AR1} we have carried out a detailed study of
the bounds obtained by considering up to three space-like constraints, both one at a time and simultaneously.
As expected, the larger the number of constraints, the smaller
is the allowed region in the $c-d$ plane.  
For one space-like constraint, data coming from the lower $|t|$ region give more stringent bounds compared to data from the higher $|t|$ region.  
On the other hand, data from higher $|t|$ allow us to include
a larger number of constraints. While this sensitivity is true for raw data from experiments, if we were to obtain theoretical fits for the data as an intermediate step it would be possible to use low $|t|$ values as well as include more constraints.

\begin{figure}
\begin{center}
\includegraphics*[angle = 0, width = 3in, clip = true]
{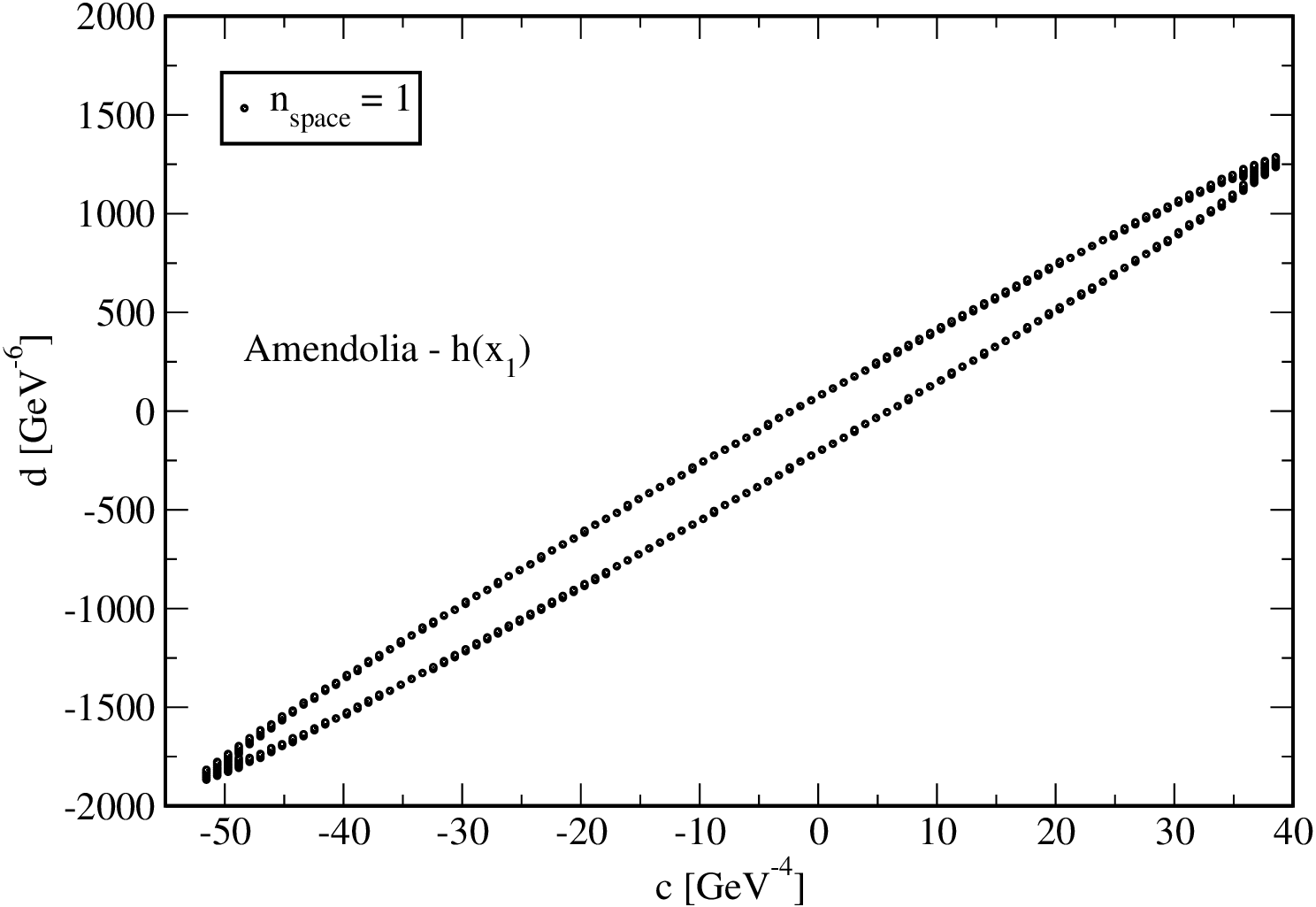}
\caption{The allowed ellipse with Amendolia data corresponding
to $x_1$.}
\label{amenfig1}
\end{center}
\end{figure}
	
\begin{figure}
\begin{center}
\includegraphics*[angle = 0, width = 3in, clip = true] 
{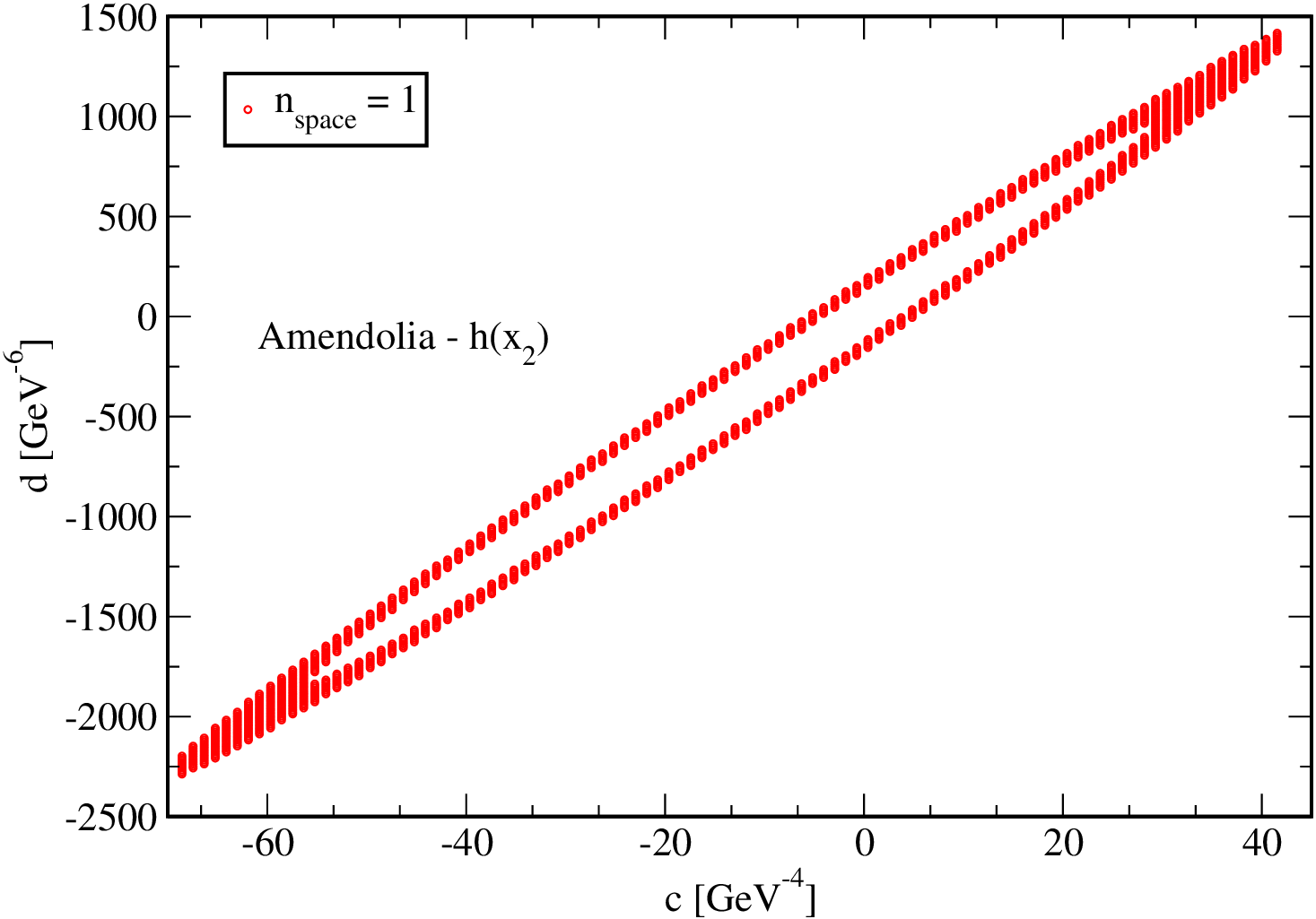}
\caption{The allowed ellipse with Amendolia data corresponding
to $x_2$.}
\label{amenfig2}
\end{center}
\end{figure}

\begin{figure}
\begin{center}
\includegraphics*[angle = 0, width = 3in, clip = true]
{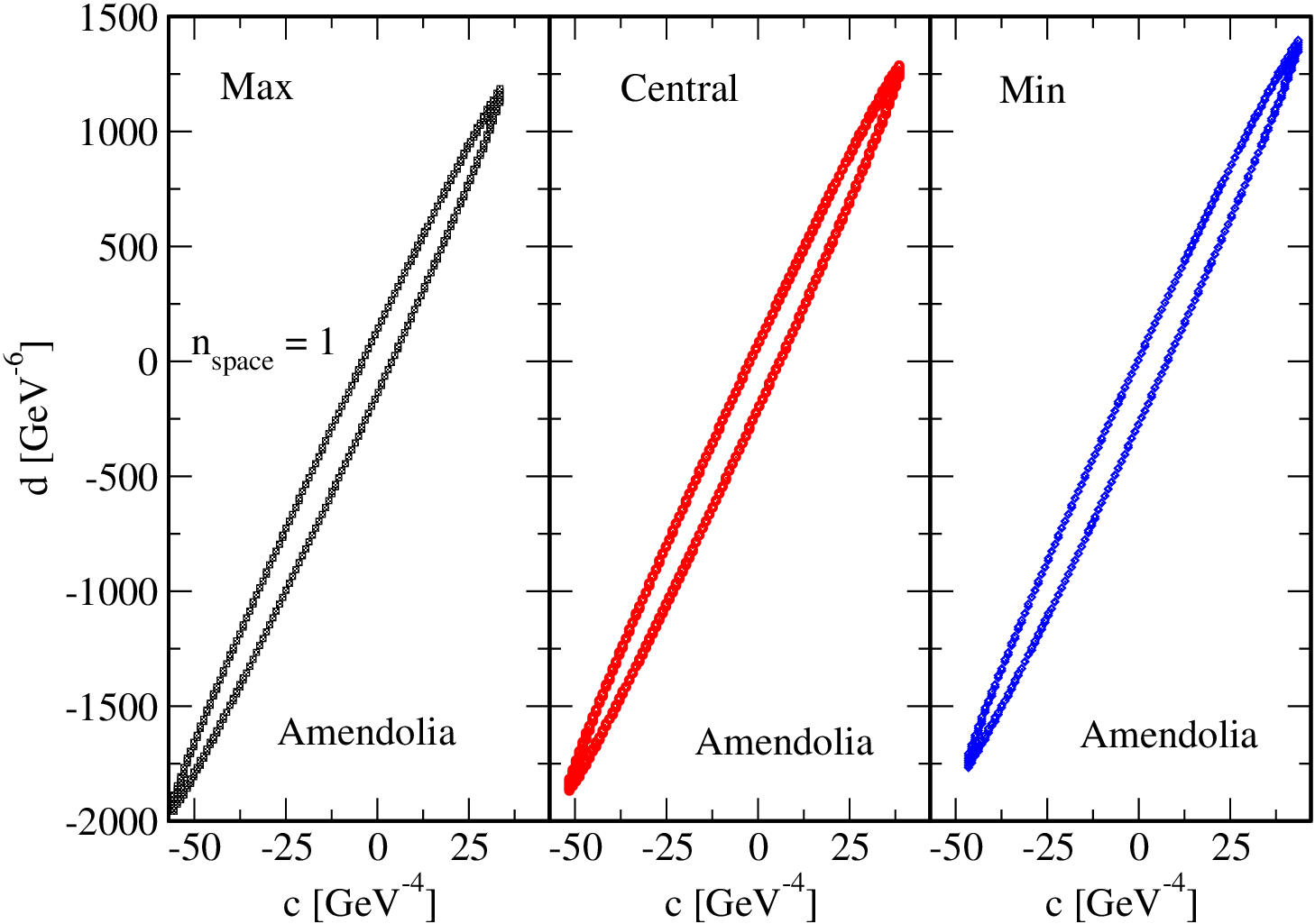}
\caption{The allowed ellipse with Amendolia data corresponding
to $x_1$, when the data is varied over the experimental uncertainty.}
\label{amenfig3}
\end{center}
\end{figure}

\begin{figure}
\begin{center}
\includegraphics*[angle = 0, width = 3in, clip = true] 
{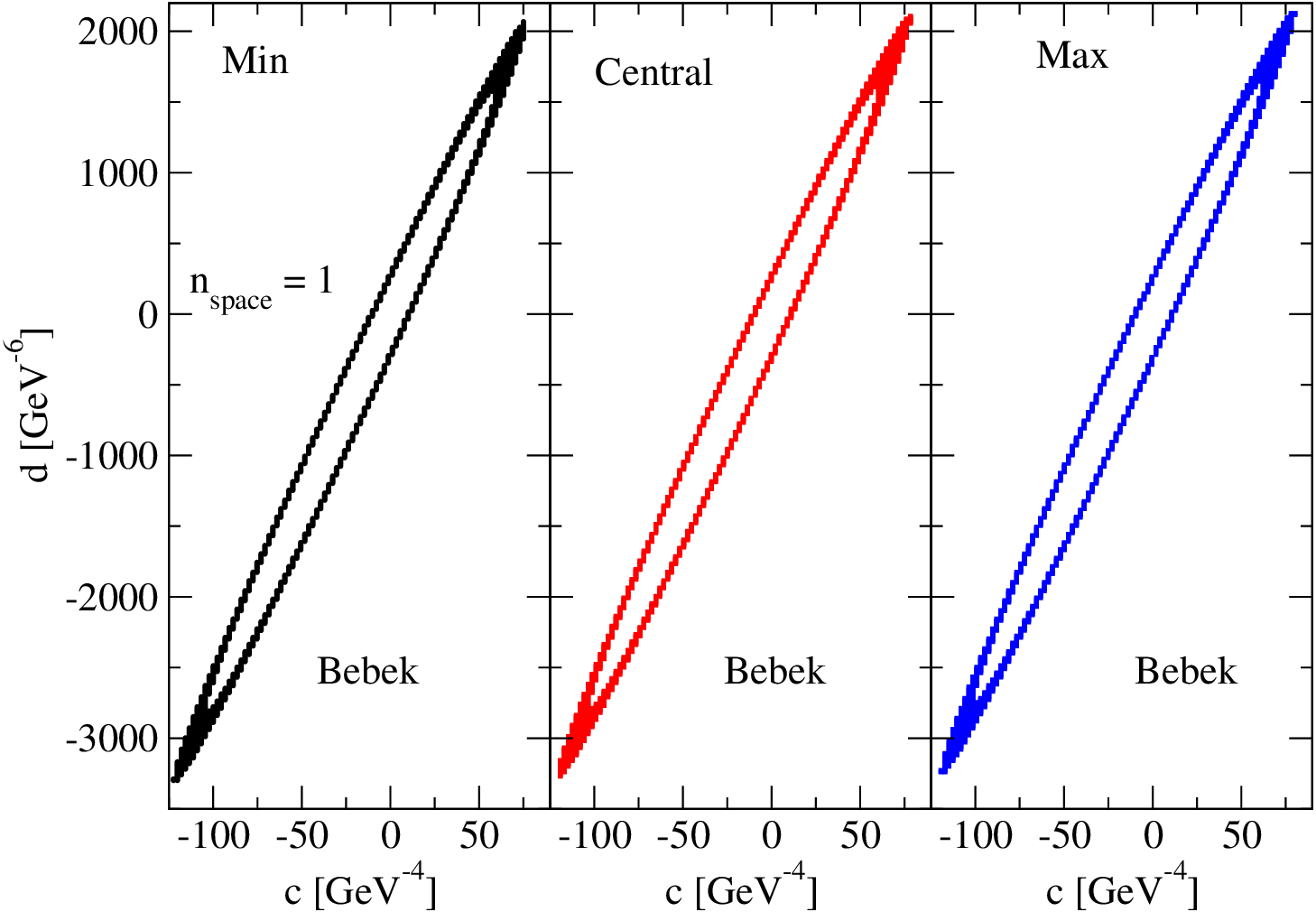}
\caption{The allowed ellipse with Bebek data corresponding
to $x_1$, when the data is varied over the experimental uncertainty. Note that the dependence on $|t|$ is rather weak as Bebek data lies in the high $|t|$ region.}
\label{Bebekvarfig}
\end{center}
\end{figure}

\begin{figure}
\begin{center}
\includegraphics*[angle = 0, width = 3in, clip = true] 
{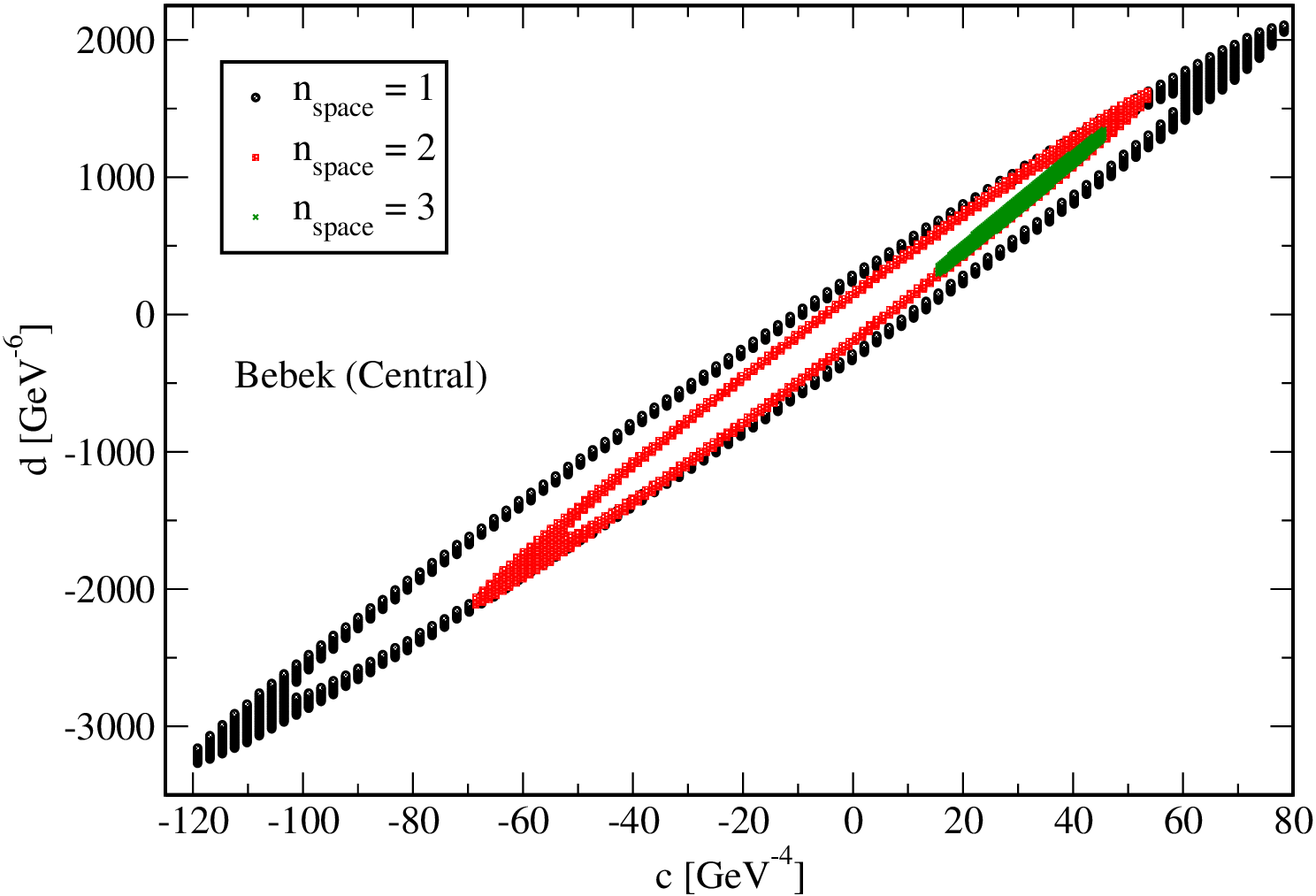}
\caption{The allowed ellipse with Bebek data corresponding
to 1 ($x_1$), 2 ($x_1,x_2$) and 3 ($x_1,x_2,x_3$)  constraints.
As the number of space-like constraints are increased, the ellipse shrinks indicating tighter bounds for $c$ and $d$.}
\label{Bebek}
\end{center}
\end{figure}

\begin{figure}
\begin{center}
\includegraphics*[angle = 0, width = 3in, clip = true]
{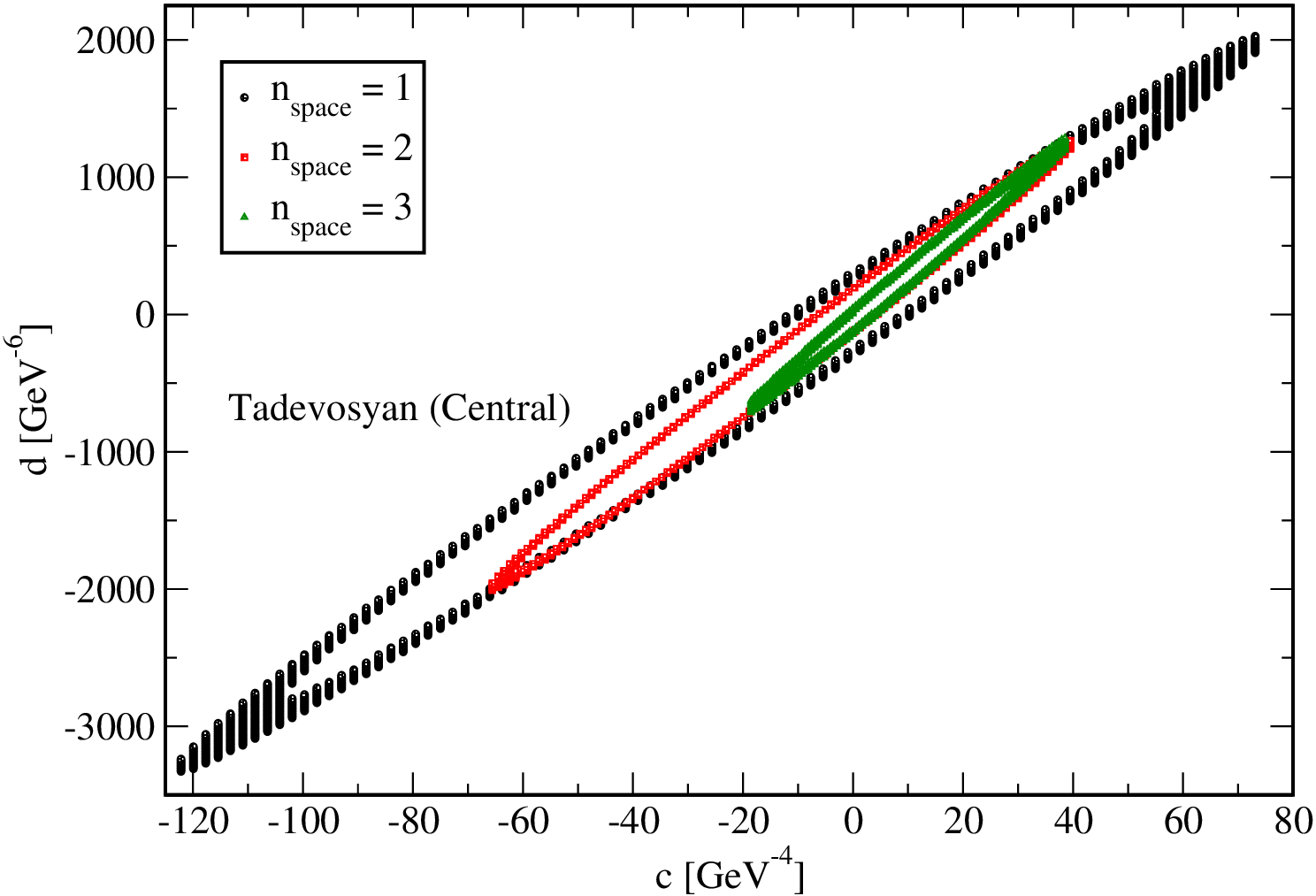}
\caption{The allowed ellipse with Tadevosyan data corresponding
to 1 ($x_1$), 2 ($x_1,x_2$) and 3 ($x_1,x_2,x_3$)  constraints. The largest one corresponds to the one space-like constraint and intermediate ellipse to two space-like constraints and the smallest to three space-like constraints.
}
\label{Tadvosyan}
\end{center}
\end{figure}

We now present the results of our analysis that are captured
in several figures.  Fig.~\ref{amenfig1} shows the allowed
region in the $c-d$ plane obtained by implementing one space-like
constraint from the Amendolia data, corresponding to the point
$x_1$.  It may be seen that the range for $c$ is now considerably
reduced to $(-55 \gev^{-4},$ $40\gev^{-4})$, whereas having no additional space-like constraints yields a range of $(-210 \gev^{-4},$ $200 \gev^{-4})$ for $c$~\cite{Caprini1}.  The
corresponding ranges with $(g-2)$ with and without additional space-like constraints are $(-20 \gev^{-4},$ $15 \gev^{-4})$ and $(-45 \gev^{-4},$ $85 \gev^{-4})$ respectively, where we see similar narrowing down of the range for $c$. 
One immediate conclusion is that the space-like data alone are
not expected to constrain $c$ significantly for either of the inputs.
Therefore, the simultaneous inclusion of space-like
constraints and the phase of time-like data, which is the subject
of the next section, becomes interesting.
In Fig.~\ref{amenfig2}, the result from the constraint with $x_2$ 
is presented.  As this corresponds to a larger value of $|t|$,
the allowed region is enlarged, as expected. In Fig.~\ref{amenfig3}
we present the results obtained with one space-like point, when the datum corresponding
to $x_1$ is varied over its experimental errors.  Some sensitivity is
seen, as the value of $|t|$ is relatively small.  In contrast,
it may be seen from Fig.~\ref{Bebekvarfig} that if a larger value
of $|t|$ is chosen, the sensitivity to the experimental error is lower.  

In Ref.~\cite{AR1} we had presented a very detailed discussion on
the bounds arising from space-like constraints alone, as these
were found to be stringent when $(g-2)$ of the muon was used as the
observable. Since the corresponding constraints with $\piprime(q^2)$ as the input are not as stringent, it is not very fruitful to investigate
them in such great detail.  Nevertheless, we show in Fig.~\ref{Bebek}
the results arising from implementing three spacelike constraints
simultaneously using the data set of Bebek et al.
It isolates a region that is at variance with
other determinations. In particular the value of $c$ favoured
by chiral perturbation theory appears to be excluded.  However, since
we are not carrying out a detailed error analysis, this result
need not be considered very significant.  The use of
three simultaneous constraints from the Tadevosyan data
leads to a result where there is no such discrepancy (see 
Fig.~\ref{Tadvosyan}). As a next step, we explore the simultaneous inclusion of the phase of time-like data with one space-like datum in the following section. The results can be easily extended to include more space-like points, which is beyond the scope of this work.

\section{Phase of time-like data and one space-like datum}
\label{sect:tl_sl}

Time-like phase is the argument of the form factor along a part of the cut. 
Consider the phase of the form factor in a region $\tpi \le t \le \tin$ in the complex $t$ plane.
Assuming that this coincides with the phase of the elastic 
two-pion scattering phase shift~\cite{Caprini1} we have, 
in accordance with the Watson final-state theorem, 
\beq
	{\rm Arg} \left[F_\pi(t+i\epsilon)\right] = 
{\delta}^1_1(t), \hspace*{0.2in} \tpi \le t \le \tin. 
	\label{phase_cap_eqn1}
\eeq
The conformal map, as defined in Eq.~\ref{ztmap.eqn1}, takes $\tpi \rightarrow z = e^{i\theta} = -1 \Rightarrow \theta = \pi$ 
and $\tin \rightarrow z = e^{i\theta} = z_{\rm in} \Rightarrow \theta = \thetain$. The upper and the lower edges of the cut in the $t$-plane is mapped onto the \emph{lower} and \emph{upper} unit disc in the $z$-plane respectively. As a result, 
\bea
	\rg {\rm Arg} \left[F_\pi(re^{i \theta})\right] &=& 
-\delta^1_1(\theta), \hspace*{0.2in} \thetain \le \theta \le \pi \\
	\rg {\rm Arg} \left[F_\pi(re^{i \theta})\right] &=& 
\delta^1_1(\theta), \hspace*{0.1in} \pi \le \theta \le (2\pi - \thetain),
	\label{phase_cap_eqn2}
\eea
where $\delta^1_1(\theta) = \delta^1_1(t(\theta))$ and $t(\theta) = t_\pi + t_\pi \text{cotg}^2(\theta)$, which is the conformal map with $z = \exp(i\theta)$.

The phase is introduced through the Omn\`{e}s function, denoted as ${\cal O}_\pi(z)$ in the $z$-plane, by,
\beq
	\rg {\cal O}_\pi(z) = \exp\left[\D\frac{i}{\pi} \int_0^{2 \pi} d\theta \frac{\bar{\delta}^1_1(\theta)}{1 - ze^{i\theta}} \right].
	\label{omnes_eqn1}
\eeq
Since the phase of the form factor along the cut (i.e. $\tpi \le t \le \tin$) is compensated by the phase of the two pion scattering phase shifts, the following condition holds,
\beq
	\rg {\rm Im} \lim_{r \rightarrow 1}\left[\D\frac{1}{{\cal O}_\pi(r e^{i \theta})} F_\pi(r e^{i \theta}) \right] = 0, 
	\label{time_like_cons1}
\eeq
which is the relevant time-like phase constraint. Substituting the expansion for $F_\pi(z)$ in terms of $h(z)$ and the outer function $w_\pi(z)$ the constraint equation becomes:
\beq
	\sum_{n = 0}^{\infty} c_n {\rm Im} \lim_{r \rightarrow 1}\left[[W(\theta)]^{-1} r^n e^{i n \theta} \right] = 0,
	\label{time_eqn1}
\eeq
where, $W(\theta \equiv \zeta = e^{i \theta}) = w_\pi(\theta) {\cal O}_\pi(\theta)$. 


Consider now the simultaneous inclusion of the above constraint (Eq.~\ref{time_eqn1}) and 
the constraint from one space-like datum:
\beq
	\rg J(z) - \sum_{n = 0}^{\infty} c_n z^n = 0.
	\label{space_eqn1}
\eeq
where, $J(z) = h(z)/\sqrt{I}$ is the space-like datum mapped on to the $z$ plane, as defined in Eq.~(\ref{ztmap.eqn1}), $I$ is the bound from $\piprime(-Q^2)$.

The space-like and time-like constraints are included through the method of Lagrange multipliers where we set up the following Lagrangian
\bea
	\lefteqn{\rg {\cal L} = \D\frac{1}{2} \sum_{n = 0}^{\infty} c_n^2} \nonumber \\ 
	&& \mbox{}+ \D\frac{1}{\pi} \sum_{n = 0}^{\infty} c_n \lim_{\rm r \rightarrow 1} \int_{\Gamma} \lambda(\theta) |W(\theta)| {\rm Im} [[W(\theta)]^{-1} r^n e^{i n \theta}] d\theta \nonumber \\
	&& \mbox{} + \alpha (J(z) - \sum_{n = 0}^{\infty} c_n z^n),
	\label{lag_eqn}
\eea
and eliminate the unknown multipliers $\alpha$ and $\lambda(\theta)$ from the Lagrange's equations and obtain an expression for the bound.	 	

The equation for $\lambda(\theta)$ was derived in Ref.~\cite{AR2} and is presented below for completeness, where we assume that the first $N$ coefficients are already constrained through the normalization of the form factor $F_\pi(0)$ and the Pion charge radius $r_\pi$. We therefore have:
\bea
	0 & = & \mbox{} -\lambda(\theta) + \sum_{n = 0}^N c_n \left[\sin(n \theta - \Phi(\theta))-\D\frac{1 - z^2}{z^{N+1}} \beta(\theta) z^n  \right] \nonumber \\
	&& \hspace*{-0.3in} \mbox{} + \D\frac{1}{\pi} \int_{\Gamma} d \thetaprime \lambda(\thetaprime) \frac{1}{2} \frac{\sin\left[(N+1/2) (\theta - \thetaprime) - \Phi(\theta) + \Phi(\thetaprime) \right]}{\sin\left[\frac{\theta - \thetaprime}{2}\right]} \nonumber \\
	&& \hspace*{-0.3in} \mbox{} + \D\frac{1}{\pi} \int_{\Gamma} d \thetaprime \lambda(\thetaprime) (1 - z^2)\beta(\theta) \beta(\thetaprime)  + J \D\frac{1 - z^2}{z^{N+1}} \beta(\theta),
	\label{lambda_eqn_main}
\eea
where 
\beq
	\beta(\theta) = \D\frac{\sin\left[(N+1) \theta - \Phi(\theta) \right] - z \sin\left[N \theta - \Phi(\theta) \right]}{1 + z^2 - 2 z \cos(\theta)}.
	\label{beta_eqn}
\eeq 
The expression for $\alpha$ is
\beq
	\alpha = \D\frac{1 - z^2}{(z^2)^{N+1}} \left[J - \sum_{n = 0}^N c_n z^n + \frac{z^{N+1}}{\pi} \int_\Gamma d\thetaprime \lambda(\thetaprime) \beta(\thetaprime) \right].
	\label{alpha_eqn2}
\eeq
We get for the following expression for the bound:
\bea
	\mu_0^2 &=& \sum_{n = 0}^N (c_n)^2 + \D\frac{1}{\pi} \sum_{n = 0}^N c_n \int_{\Gamma} d \theta \lambda(\theta) \sin\left[n \theta - \Phi(\theta) \right] \nonumber \\
	&& \mbox{} + \alpha \left(J - \sum_{n = 0}^N c_n z^n \right) \le 1.
	\label{mu_eqn2}
\eea 
In our case, we will set $N= 3$ following~\cite{Caprini1}, 
where $\Gamma={\theta:\theta_{in}<\theta<2\pi-\theta_{in}}$.
We solve for $\lambda(\theta)$ using Eq.~(\ref{lambda_eqn_main}), 
obtain the corresponding value of $\alpha$ from Eq.~(\ref{alpha_eqn2}). 
Using $c_n$'s,  already constrained by normalization of $F_\pi$ and pion 
charge radius, $\mu_0^2$ is evaluated (Eq.~(\ref{mu_eqn2})). 
Only those coefficients, $c_n$, which satisfy $\mu_0^2 < 1$ are retained. Our general formalism has been cross-checked against related results
presented in~\cite{BC} for the pi K system with $N=2$. 

The space-like datum we use for our results are the data points corresponding to $t(-Q^2)$ $ =$ $-0.6$ $ \gev^2$ from the Tadevosyan data set (Table~\ref{table_jlab}) and $t(-Q^2)$ $ =$ $-0.131$ $\gev^2$ from the Amendolia data set (Table~\ref{table_amen}). They are referred to as ``Tadevosyan'' and ``Amendolia'' respectively in the text and figures. 

The time-like phase is defined as~\cite{Caprini1},
\beq
	\delta^1_1(t) = {\rm arc}\tan \left(\D\frac{m_\rho \Gamma_\rho(t)}{m_\rho^2 - t}\right),
	\label{delta_eqn1}
\eeq
and
\beq
	\Gamma_\rho(t) = \D\frac{m_\rho t}{96 \pi f_\pi^2} \left(1 - \frac{4 m_\pi^2}{t} \right)^{3/2},
\eeq	
where $m_\rho = 770 \mev$ is the mass of the $\rho$ meson, 
$\Gamma_\rho = 150 \mev$ is the width of the $\rho$ resonance 
and $m_\pi = 139 \mev$ is the mass of the Pion. At low energies 
Eq.~(\ref{delta_eqn1}) agrees well with the one-loop chiral 
perturbation theory expression for the two-pion elastic scattering 
phase shifts and also with experiments for $t \ge 0.5 \gev^2$, 
as noted in~\cite{Caprini1} (see also
Ref.~\cite{hep-ph/9707347}).  Therefore, we assume that the phase of 
the pion form factor coincides with Eq.~(\ref{delta_eqn1}) 
for $\tpi < t < \tin$, where $\tin = 0.8 \gev^2$. 
We also use the results for the phase shift coming from
the Roy equation analysis of Ref.~\cite{ACGL} given in 
Eqn.~(D.1) therein, for two choices of input parameters that
are $(a^0_0,a^2_0)=(0.225,-0.0371)$ (ACGL) and 
$(0.220,-0.0444)$\cite{CGL} (CGL). For most of the work, we have chosen $t_{in}$ to be $0.8\gev^2$
in order to have a meaningful comparison with the work of Caprini.

\begin{figure}
\begin{center}
\includegraphics*[angle = 0, width = 3in, clip = true]
{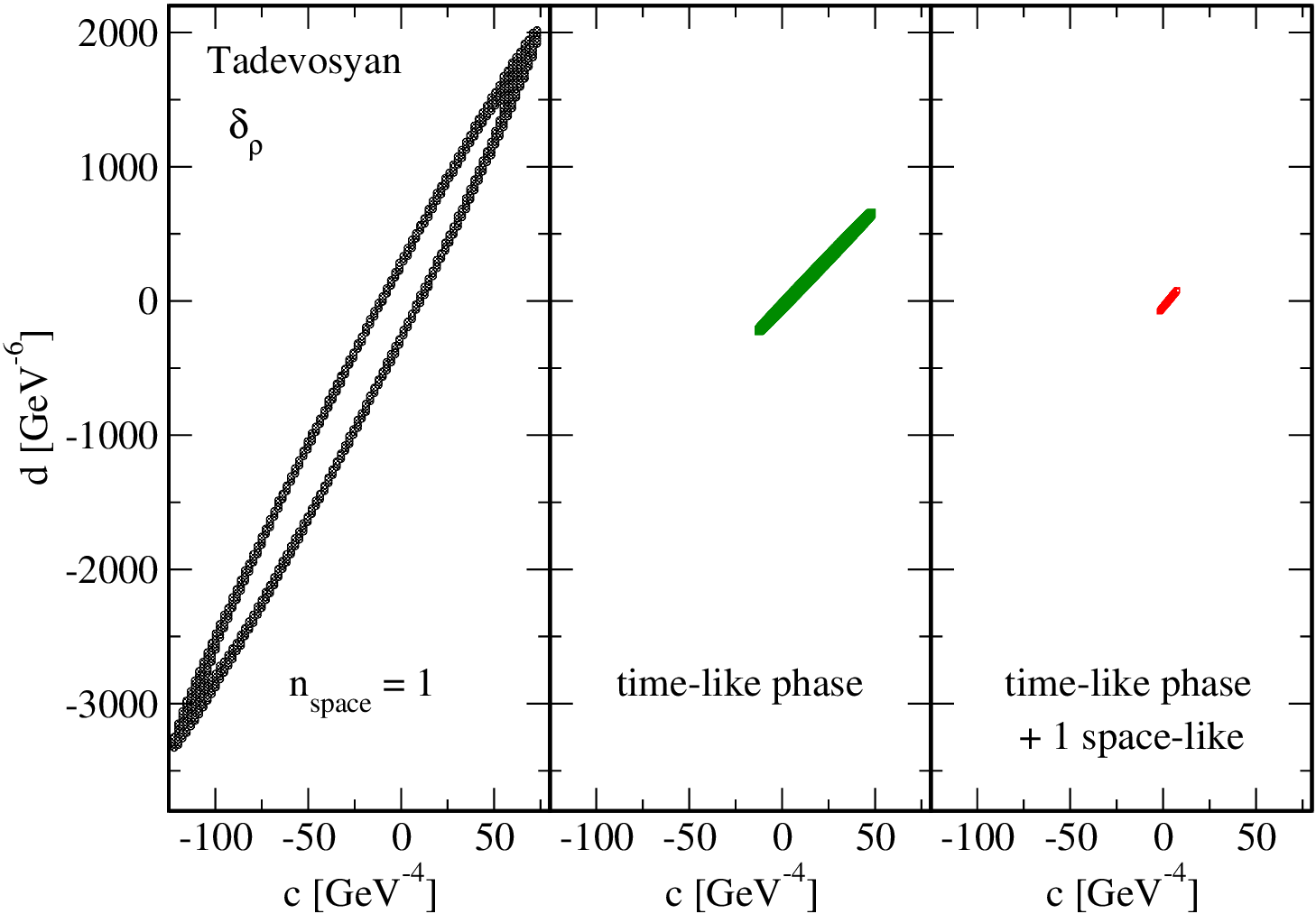}
\caption{Bounds on the expansion coefficients $c$ and $d$. Left panel shows the bounds obtained from one pure space-like datum, the middle panel from the pure phase of time-like data and the right from the phase of time-like and one space-like datum taken together. Note that the simultaneous inclusion substantially
shrinks the allowed ellipse.}  
\label{sl_tad1}
\end{center}
\end{figure}

\begin{figure}
\begin{center}
\includegraphics*[angle = 0, width = 3in, clip = true]
{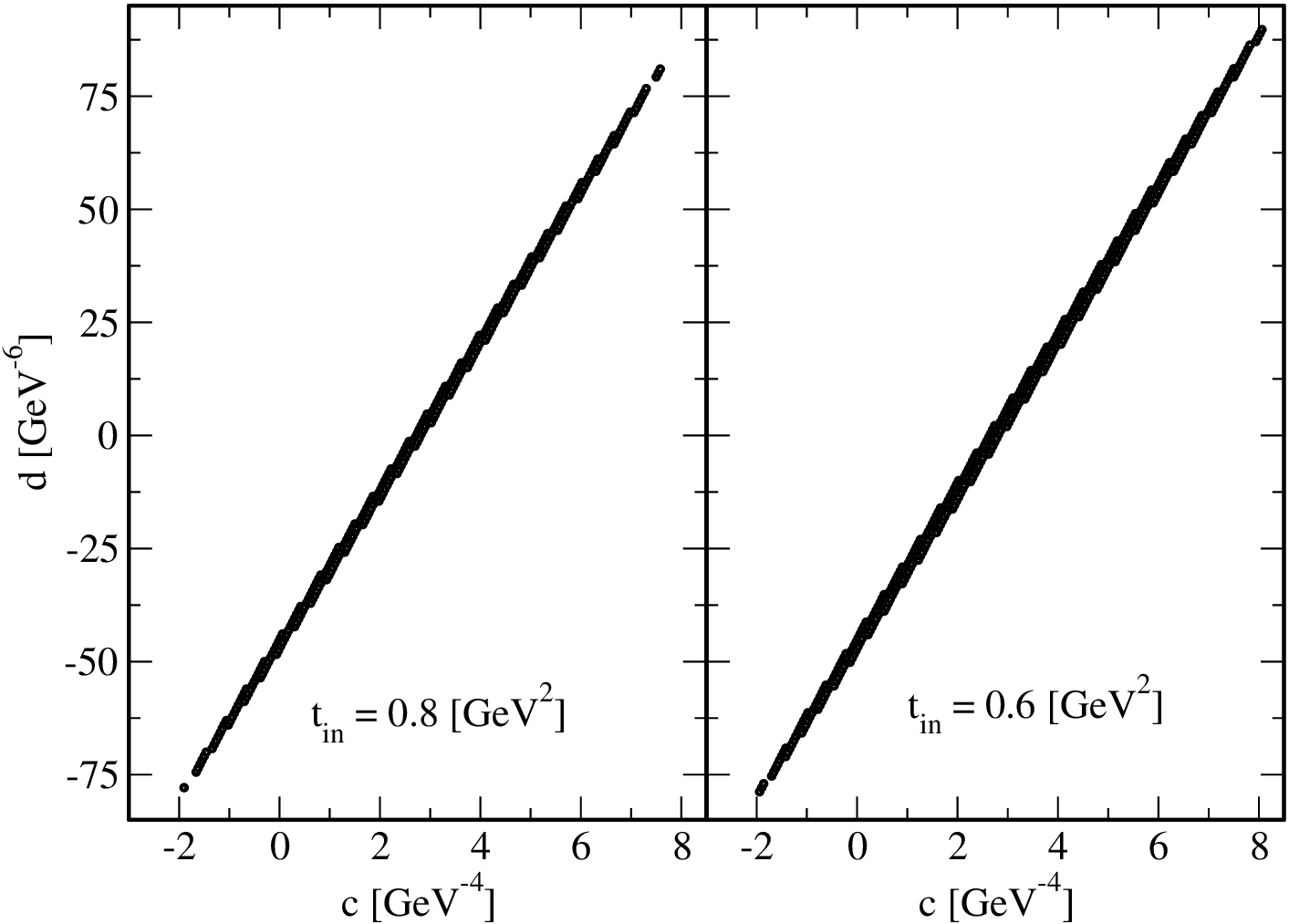}
\caption{The allowed ellipse with Tadevosyan datum and phase
of time-like data with $\tin = 0.8 \gev^2$ (left panel) and $\tin = 0.6 \gev^2$ (right panel). Note the weak dependence on $\tin$.}  
\label{sl_tad2}
\end{center}
\end{figure}

\begin{figure}
\begin{center}
\includegraphics*[angle = 0, width = 3in, clip = true]
{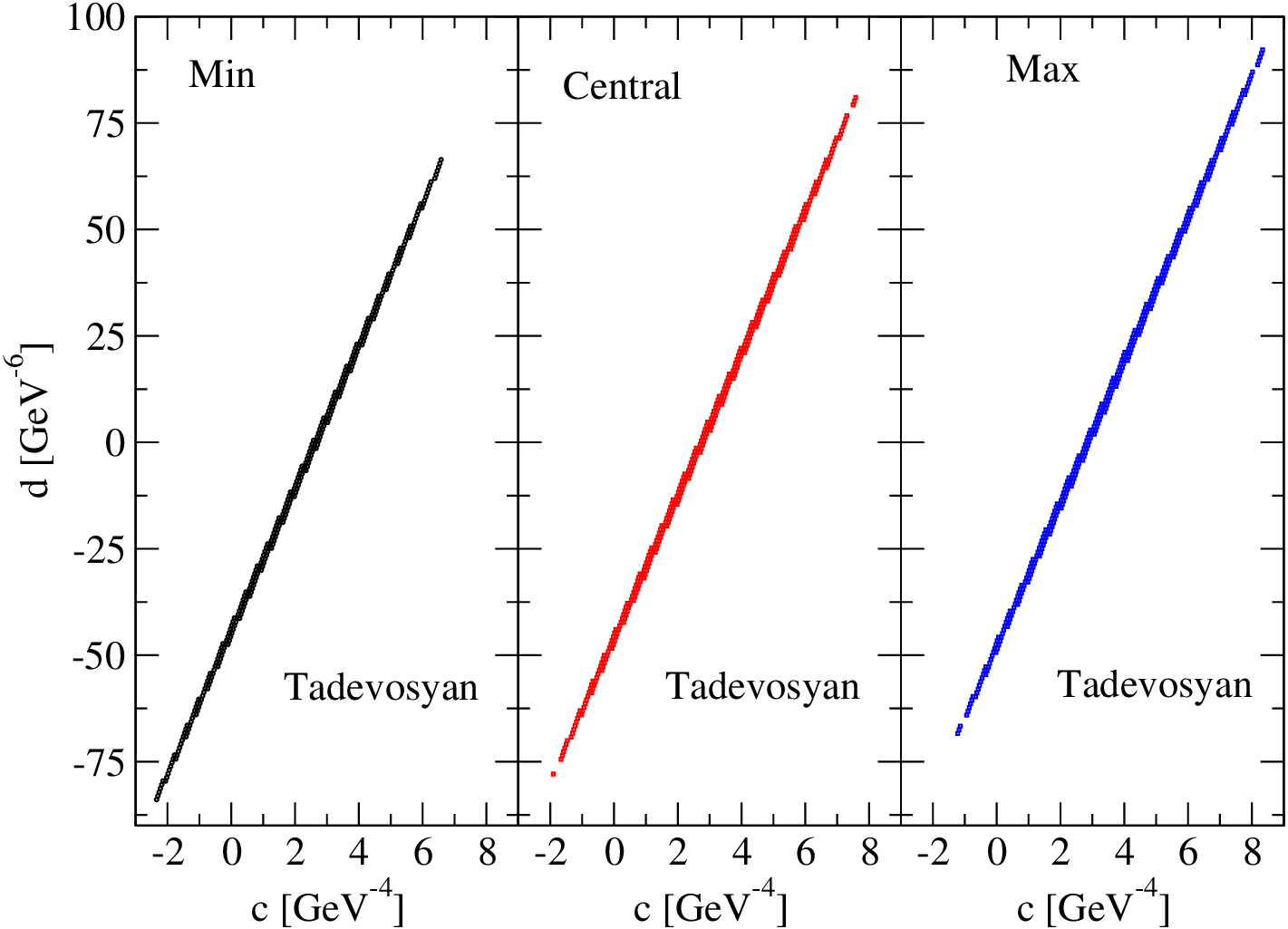}
\caption{The allowed ellipse with Tadevosyan datum and phase
of time-like data, when the datum is varied over its experimental
uncertainty}  
\label{tadcmm}
\end{center}
\end{figure}

\begin{figure}
\begin{center}
\includegraphics*[angle = 0, width = 3in, clip = true]
{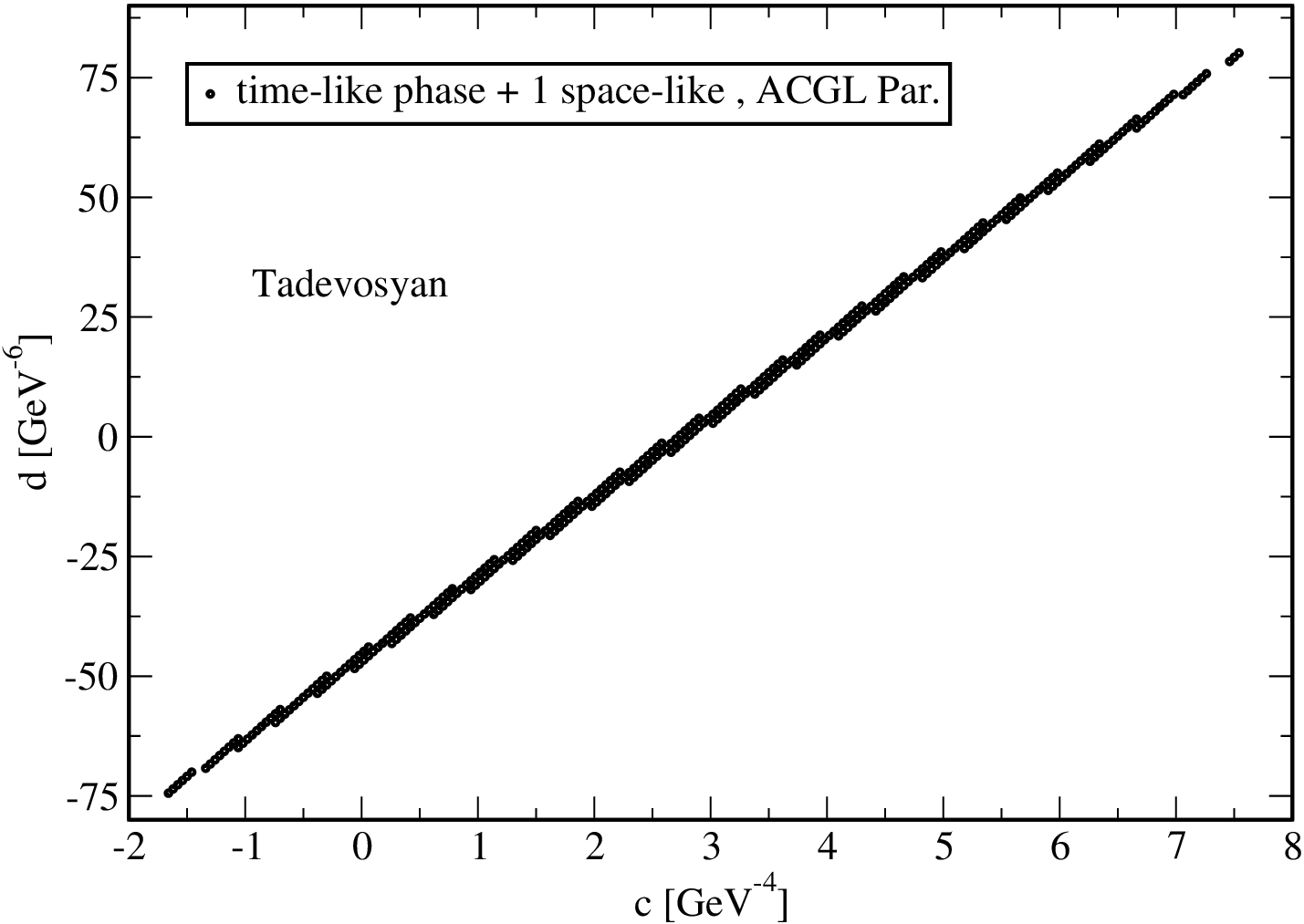}
\caption{The allowed ellipse with Tadevosyan datum and phase
of time-like data from ACGL~\cite{ACGL}}  
\label{tadacgl}
\end{center}
\end{figure}
\begin{figure}
\begin{center}
\includegraphics*[angle = 0, width = 3in, clip = true]
{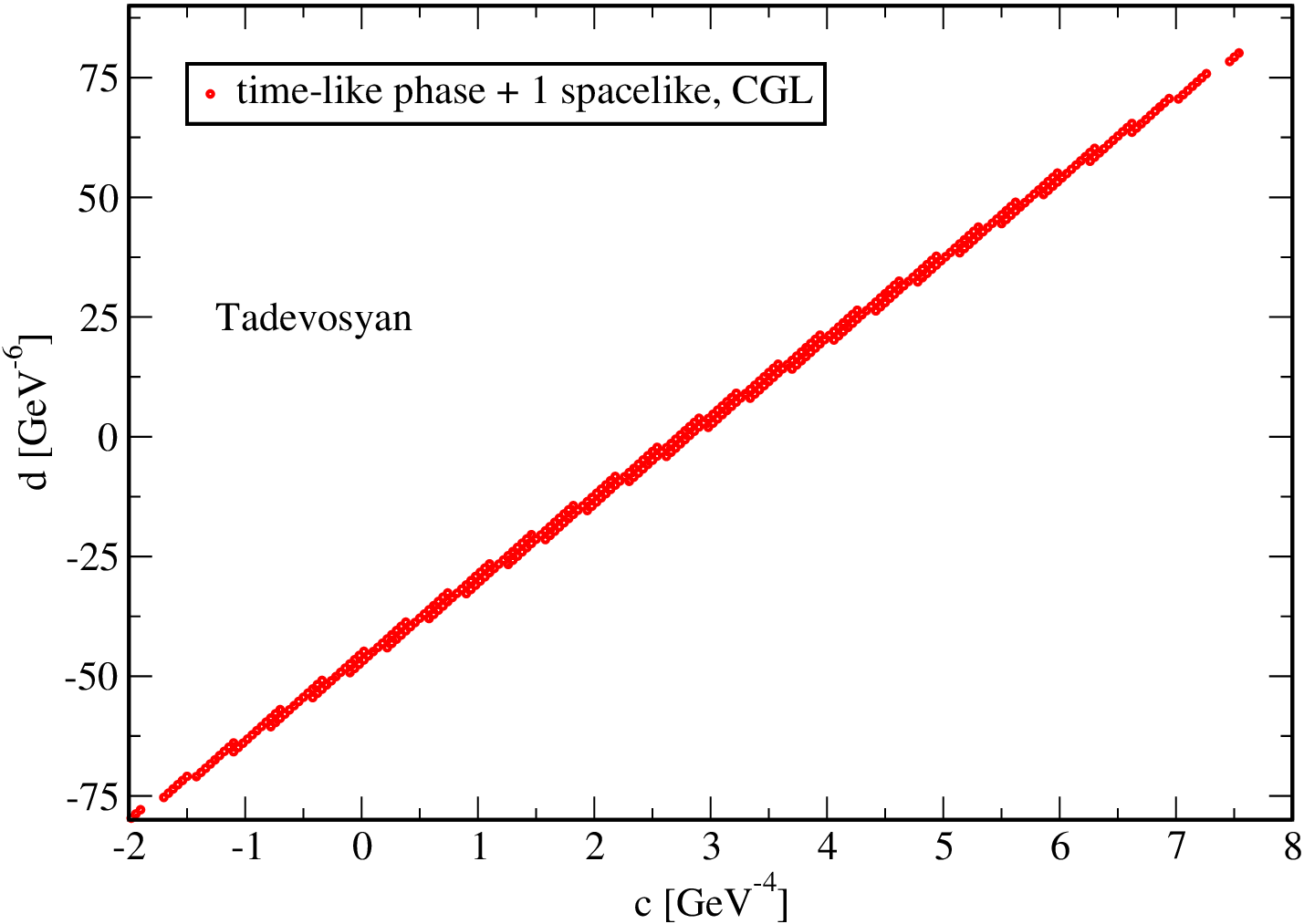}
\caption{The allowed ellipse with Tadevosyan datum and phase
of time-like data from CGL~\cite{CGL}}  
\label{tadleut}
\end{center}
\end{figure}

In Fig.~\ref{sl_tad1} we present the bounds when the phase of the time-like data with
the analytical model for the phase shift (Eq.~\ref{delta_eqn1}) and the space-like
datum from Tadevosyan.
The weak bounds obtained from the space-like data (left panel) is already improved by the inclusion of time-like phase alone (central panel). When the data from the space-like and the time-like region are used together, the allowed ellipse shrinks significantly as seen in the right panel in Fig.~\ref{sl_tad1}.
Note that the
pure time-like phase constraint ellipse shown in this figure was
already presented in Ref.~\cite{Caprini1}, while the bounds from pure
space-like datum is shown in Fig.~\ref{Tadvosyan}.
In Fig.~\ref{sl_tad2} we present a close up view of the
allowed region for $\tin = 0.8 \gev^2$ as well as for $\tin = 0.6 \gev^2$. We observe that the bounds do exhibit some sensitivity to the choice of $\tin$. In order to test the sensitivity of
the results to other inputs, we now vary the space-like datum over its allowed experimental
range. The result is presented in Fig.~\ref{tadcmm}. In order to judge the dependence on the parametrization of
the phase shift, we use the two Roy equation fits ACGL and
CGL~\cite{ACGL,CGL}. The results for these are shown in Figs.~\ref{tadacgl} and~\ref{tadleut}.
From Figs.~\ref{tadcmm},~\ref{tadacgl} and~\ref{tadleut}, we can conclude that the bounds are sensitive to the errors in the space-like data, while they are not significantly affected by the parametrization for the phase shifts.

In order to study the dependence on $|t|$ we consider the datum
from the Amendolia data. The bounds (not shown here) are very well constrained, however, such small $|t|$ data lead to results
that may be highly sensitive to the errors in the experiment and consequently are unreliable unless we use a theoretical fit for the data, which is beyond the scope of our current work. Hence we do not display these results.

\begin{figure}[ht]
 	\begin{center}
 	 \includegraphics*[angle = 0, width = 3in, clip = true]{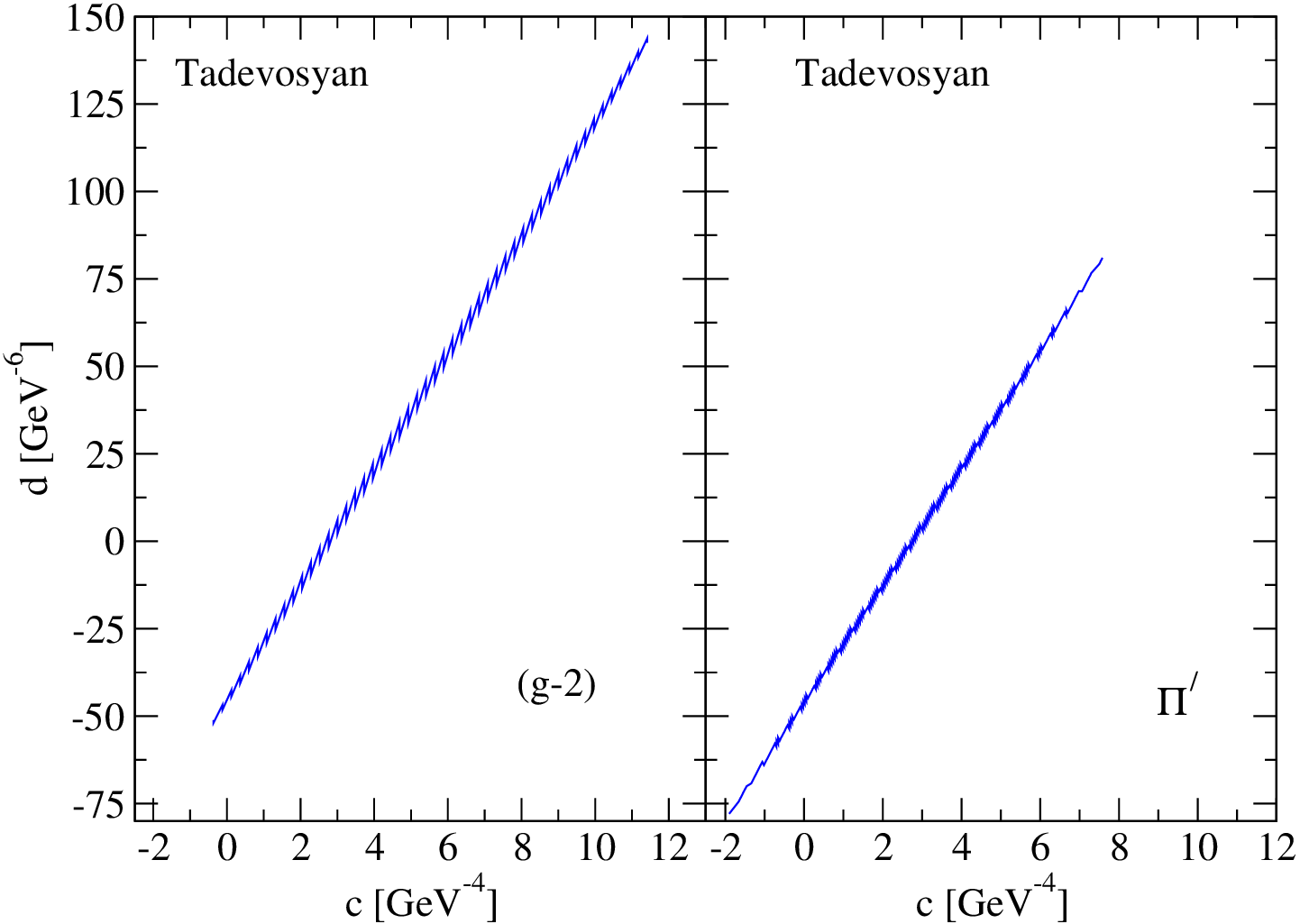}
	\caption{Comparing the results for the bounds on $c$ and $d$ using $a_\mu$ (left panel)~\cite{AR2} and $\piprime(q^2)$ (right panel) as inputs. The space-like datum comes from the set of Tadevosyan in Table~\ref{table_jlab}, while the time-like phase uses the model given in Eq.~\ref{delta_eqn1}}.
	\label{amu_piprime.fig}
 	\end{center}
\end{figure}

It may be concluded from this analysis, that the simultaneous
inclusion of the phase of time-like data and one space-like datum
leads to values of $c$ in the range $(-2$ $\gev^{-4},$ $8$ $\gev^{-4})$ while the
range for $d$ is in the range $(-75$ $\gev^{-6},$ $80$ $\gev^{-6})$.  In particular from Fig.~\ref{amu_piprime.fig},
it is observed that the latter is significantly more constrained when $\piprime(q^2)$ is used as the input instead of $(g-2)$ of the muon.  The ranges of $c$ in both the cases have
a considerable overlap, which safely accommodates the various
prior determinations of these quantities in chiral perturbation
theory. It is worth emphasizing that the interplay of space-like and phase of time-like constraints together have narrowed down $c$ and $d$ to an extent that one would not have anticipated from the results of either taken separately for $\piprime$.  

\section{Discussion and Summary}
\label{sect:conclusion}

In this paper, we study the improvements on the bounds of the low-energy 
Taylor expansion coefficients of the Pion EM form factor, when both phase 
of time-like data and one space-like datum are used. 
We use the method of Lagrange multipliers to include the constraints.
In contrast to Ref.~\cite{AR1,AR2} we have used 
$\piprime(-Q^2)$ as an input.  This was used by Caprini~\cite{Caprini1}, but without
the use of space-like constraints.  

We have found that many of the earlier 
determinations~\cite{AR1,AR2,Caprini1} are reinforced by the present
work, in particular the determination of the allowed region for
$c$.  However, each of these studies provides a different allowed
region in the $c-d$ plane, although there are regions of intersection.
As in~\cite{AR2}, we can try and inspect the region of intersection of the ellipse in e.g. Fig.~\ref{sl_tad2} with that of Fig.~3 of~\cite{Caprini1} that uses the phase as well as the modulus of the time-like data along a part of the cut. It appears that for $c \sim 4.5 \gev^{-4}$, there is no longer an overlap. However, for smaller values such as $c \sim 4.0 \gev^{-4}$ and ever more so for $c = 3.5 \gev^{-4}$ there is substantial overlap. 
However, a complete picture can be arrived at only when the latter
analysis is subjected to tests on the sensitivity of the results
to variations in $t_{in}$ as well as model dependence.  

We find it heartening that so many different approaches have led to a coherent
picture for these important expansion coefficients of the pion
electromagnetic form factor.  In particular, the value of $d$ is typically about
$10 \gev^{-6}$ for $c \sim 3.2 \gev^{-4}$ and about $20 \gev^{-6}$ for $c \sim 4.0 \gev^{-4}$. The value of $d$ for $c \sim 3.2 \gev^{-4}$ agrees with the that from~\cite{Caprini1} using constraints from phase as well as modulus of the form factor. In~\cite{AR2}, we pointed out that the phenomenological fits to the form factors from ALEPH also yield a number for $d$ $\sim$ $10\gev^{-6}$ that agrees with our determination at lower $c$ values. Another discussion may be found in~\cite{Truong_98}, where $(c,d)$ = $(3.9$ $\pm$ $0.1$ $\gev^{-4}$, $9.7$ $\pm$ $0.4 \gev^{-6})$ are reported. 

Our results for the bounds on the Taylor coefficients using time-like phase as well as one space-like datum can be extended to include the modulus of the form factor along a part of the cut. Combined error analysis of space-like and time-like data will yield predictions for $c$ and $d$ with reliable error estimates, which could be considered in the future.

\begin{acknowledgement}
BA thanks Department of Science and Technology, Government of India
for support. SR thanks the Abdus Salam International Centre for Theoretical
Physics, Trieste, Italy, for its hospitality when part of this work
was done. 
\end{acknowledgement}


\end{document}